\def\rmb{{\rm b}}

\def\rmd{{\rm d}}
\def\rme{{\rm e}}

\def\rmp{{\rm p}}
\def\rmt{{\rm t}}

\def\rmK{{\rm K}}
\def\rmT{{\rm T}}

\def\bfv{{\bf v}}

\def\kms{\, \rm{km\,  s^{-1}}}

\def\etal{{et al. }}
\def\HH{${\rm {H_2}}\,\,$}

\def\cm{{\rm cm}}

\def\etal{{et al. }}
\def\HH{${\rm {H_2}}\,\,$}

\def\cm{{\rm cm}}

\def\gs{\mathrel{\raise1.16pt\hbox{$>$}\kern-7.0pt
\lower3.06pt\hbox{{$\scriptstyle \sim$}}}}
\def\ls{\mathrel{\raise1.16pt\hbox{$<$}\kern-7.0pt
\lower3.06pt\hbox{{$\scriptstyle \sim$}}}}

\def\gtsima{$\; \buildrel > \over \sim \;$}
\def\ltsima{$\; \buildrel < \over \sim \;$}
\def\prosima{$\; \buildrel \propto \over \sim \;$}
\def\gsim{\lower.5ex\hbox{\gtsima}}
\def\lsim{\lower.5ex\hbox{\ltsima}}
\def\simgt{\lower.5ex\hbox{\gtsima}}
\def\simlt{\lower.5ex\hbox{\ltsima}}
\def\simpr{\lower.5ex\hbox{\prosima}}

\def\pp{\noindent\parshape 2 0truecm 17truecm 2truecm 15truecm}
\def\rf#1;#2;#3;#4 {\par\pp#1, #2, #3, #4. \par}

\def\pr{\ref@jnl{Phys.Rev}}

\newcommand{\href}[2]{{\bf #2} (\texttt{#1})}

\def\ssubsection#1 {\subsection{\sl #1}}

\def\beq#1{\begin{equation}\label{#1}}
\def\eeq{\end{equation}}
\def\beqa#1{\begin{eqnarray}\label{#1}}
\def\eeqa{\end{eqnarray}}
\def\eq#1{equation~(\ref{#1})}
\def\Eq#1{Equation~(\ref{#1})}

\def\tento#1{\times 10^{#1}}

\def\K{{\rm \ K}}
\def\s{{\rm \ s}}

\def\kms{{\rm \ km/s}}

\def\cm{{\rm \ cm }}
\def\eV{{\rm \ eV }}

\def\Mpc{{\rm \ Mpc }}
\def\kpc{{\rm \ kpc }}
\def\pc{{\rm \ pc\ }}

\def\yrs{{\rm \ years }}

\def\HH{H$_2$ }
\def\H2p{H$_2^+$ }
\def\HHp{H$_2^+$ }

\def\Hm{H$^-$ }

\def\mH2p{H_2^+}

\documentstyle[epsfig,  aj_pt4]{article}

\begin{document}

\thispagestyle{empty}

\title{{\huge \bf \sc First Structure Formation: \\ \vspace{.2cm}
  II. Cosmic String + Hot Dark Matter Models.}}

\author{\bf{Tom Abel$^{1,2}$, Albert Stebbins$^3$, Peter Anninos$^1$, \&
  Michael L. Norman$^1$}  \\
\  \\ }
\vspace{2cm}
\affil{\flushleft \small
\texttt{abel@mpa-garching.mpg.de, stebbins@fnal.gov,  \\
 panninos@ncsa.uiuc.edu, norman@ncsa.uiuc.edu} \\
 \ \\
{\it $^1$ Laboratory for Computational Astrophysics}       \\
{\it      National Center for Supercomputing Applications} \\
{\it      University of Illinois at Urbana--Champaign}     \\
{\it      405 N. Mathews Ave., Urbana, IL, 61801},         \\
\ \\
{\it $^2$ Max-Planck-Institut f\"ur Astrophysik}           \\
{\it      85748 Garching, Germany},                        \\
\ \\
{\it $^3$ NASA/Fermilab Astrophysics Center}               \\
{\it      Fermilab, Box 500, Batavia, Ill, 60510-0500} \flushright
\rm \today}
\vspace{1cm}

\begin{abstract}
  We examine  the  structure of  baryonic wakes in  the
  cosmological fluid which would  form behind GUT-scale cosmic strings
  at early times (redshifts $z\gsim 100$)
  in an neutrino-dominated universe.  We
  show,  using  simple    analytical arguments   as    well as  1-  and
  2-dimensional hydrodynamical simulations,  that these wakes will {\em
  not} be able  to  form interesting  cosmological objects before  the
  neutrino  component  collapses.  The  width  of the  baryonic  wakes
  ($\lsim10$\,kpc  comoving) is smaller  than the  scale of wiggles on
  the strings and  are probably not enhanced  by the wiggliness of the
  string network.
\end{abstract}

\section{INTRODUCTION}
\label{sec:introduction}

One of the  more interesting  unknowns  in cosmology  is what happened
during  the ``dark   ages'',  the  period of   time   between the last
scattering of the  background radiation photons ($z\sim10^3$), and the
formation  of the first   objects  we see,   QSO's  ($z\sim  5$).   As
observational techniques improve, we   are able to examine objects  in
the universe at earlier and earlier cosmological  epochs (cf.  Wampler
\etal 1996,  Abraham \etal 1996,   and references therein).  There  is
strong   evidence that objects the     size of very large  present-day
galaxies  existed at  $z\sim4$.  The  plethora  of early objects found
suggest that the dark ages were not completely quiescent.  However the
dark ages could  not have been too active  a period since the spectrum
of the  Cosmic Microwave Background  Radiation (CMBR), being extremely
close to  a blackbody  spectrum,  precludes the   presence of a  large
amount of hot gas ($\gg10^3$K) at  these early times.  This leaves the
possibility   of  small sub-galactic   objects,   with   small  virial
temperatures, being formed at very early times, $z\gg5$.

        Most inflationary models for  the origin of inhomogeneities in
the universe   predict that  the   primordial density  field will   be
Gaussian       random  noise      with   a   nearly    scale-invariant
Harrison-Zel'dovich spectrum.  The shape of the spectrum combined with
the Gaussian  statistics  cause object  formation  to  turn on  fairly
rapidly with very few precursor objects. In contrast, the non-Gaussian
nature of inhomogeneities seeded  by  topological defects allow  for a
much greater number of  precursor objects of a  given mass to collapse
long  before  the  rms  fluctuation    on that  mass   scale has  gone
non-linear.  Such  precursor  objects may play  an  important  role in
early star formation, producing metals  to increase the efficiency  of
cooling in subsequent objects, and perhaps even triggering star-bursts
via the shock waves produced by supernovae.

        In  the cosmic string model,  these precursor objects can form
around the  wakes of cosmic  strings which move through the primordial
gas at all   epochs (see Vilenkin and   Shellard 1994 for a  review of
cosmic strings).   It has  been argued that   star formation can  take
place in these  wakes as early  as $z\sim10^2$ (Rees 1986,  Hara 1987,
Hara 1996a) and that  one  might very well form   black holes as  well
(Hogan 1984, Hara \etal 1996b).  However the analysis which has led to
these conclusions  has not   been sufficiently detailed   to determine
convincingly how, when, and whether these events occur.  In this paper
we examine  some of these   scenarios with hydrodynamical simulations.
In particular we  look at the most easily   posed problem of a  single
cosmic string wake in a flat universe with  hot dark matter (HDM).  We
emphasize at the outset that our conclusions  do not apply to wakes in
a universe with cold dark mater (CDM), or one with a mixture (MDM), or
a universe without non-baryonic  dark matter (so  called BDM).  We are
also  only considering the wakes {\it  before}  the neutrinos begin to
collapse.  Here we   do  not consider the  collapse  of  the neutrinos
themselves which will happen after the  events we describe take place.
There  is  ample literature  on this   subject  (Stebbins \etal  1987;
Brandenberger, Kaiser, \&  Turok 1987;  Bertschinger \& Watts,   1988;
Scherrer,     Melott,     \&  Bertschinger   1989;   Perivolaropolous,
Brandenberger, \& Stebbins  1990; Colombi   1993; Zanchin, Lima,    \&
Brandenberger 1996, Moessner and Brandenberger 1997, Sornborger
1997). 

        Of  course   it is   difficult   to  demonstrate,  using   the
hydrodynamical simulation techniques   available today, the nature  of
star  formation   in a given  cosmological  scenario.   However we are
attempting the easier task  of  determining where  stars {\it do  not}
form.  In the cosmic string  + HDM scenario, just after recombination,
the gas  is smoothly distributed since  small  scale perturbations are
damped by photon diffusion (Silk  damping) while  the HDM is  smoothly
distributed because of it's  large velocity dispersion  (i.e.  damping
due to free-streaming).  Thus in this  scenario we can hope to resolve
all relevant scales in the problem, or at least before things collapse
too  much.  Clearly stars cannot form  unless the gas density of grows
enormously over the ambient density.  One  might expect this to happen
due to a cooling instability.  However if  the gas does not undergo an
instability leading to very large over-densities then we can be fairly
certain that  stars  will not     form  in that  region.  With     the
comprehensive non--equilibrium cooling  and  chemistry  model of  Abel
\etal 1997a, our study   implements the details of  molecular hydrogen
formation and   cooling accurately.  Since  our results  indicate that
instabilities generally do not  occur, we do  not need to  worry about
secondary effects which may be produced by star formation.

To study this problem we have  used a version  of ZEUS--2D, a Eulerian
finite difference hydrodynamics  code,  that was modified  to simulate
nonequilibrium  reactive flows in   cosmological sheets by Anninos  \&
Norman (1996).  The code incorporates the methods developed by Anninos
\etal (1997), and the comprehensive  9-species chemistry model of Abel
\etal  (1997). For this  study we  have further  modified  the code to
account  for  the  net  force imparted  on  the baryons   from Compton
scattering with the CMBR), the
so called {\it Compton drag}.

In the   next section we    give analytical  arguments why  a  cooling
instability is not found  in the cosmic string +  HDM model.   Then we
will    describe our  numerical     methods  and results in    section
\ref{sec:simulations}.   We    refer the  interested    reader  to our
\href{http://lca.ncsa.uiuc.edu/$\sim$tom/Strings/}{WWW page} that also
shows visualizations  and movies   of the 2--D  simulations.   Section
\ref{sec:discussion}    gives a brief   summary    of our results  and
discusses the implications for the cosmic string + HDM model.

\section{ANALYTICAL ARGUMENTS}
\label{sec:AModel}

In this section we  present  some simple  analytical arguments on  the
evolution  of  baryonic  wakes.  The  velocity perturbations in the
primordial gas  induced by a moving string  will cause  the baryons to
shock heat at the position  of the string's  world sheet. For a ``planar
string'' the velocity perturbations show  reflection symmetry along the
world sheet. Hence the evolution of  the baryons will show the
same      symmetry   unless  hydrodynamic   or thermo--gravitational
instabilities affect the flow.

Throughout this paper we assume a hot dark matter component that fills
the universe with $\Omega_{HDM} + \Omega_B =1$. The velocity
dispersion of particles that were thermally produced in the early
universe is given by (Kolb \& Turner, 1990)
\begin{eqnarray}
      \langle   v_{HDM}(z)  \rangle \approx 23 \kms \left( \frac{m_{\nu}}
        {30\eV} \right)^{- \frac{4} {3}} \left( \frac{1 + z}{10} \right),
\end{eqnarray}
where $m_{\nu}$ denotes the mass of the HDM particle. Since we are
interested in early structure formation ($z\gg~5$), we can safely
assume that the HDM component is distributed homogeneously and
is not therefore dynamically important.
Hence, $\Omega_{HDM}$ will only enter in the evaluation of
the cosmological scale factor $a(t)$.

\begin{figure}[ht]
\centerline{\psfig{file=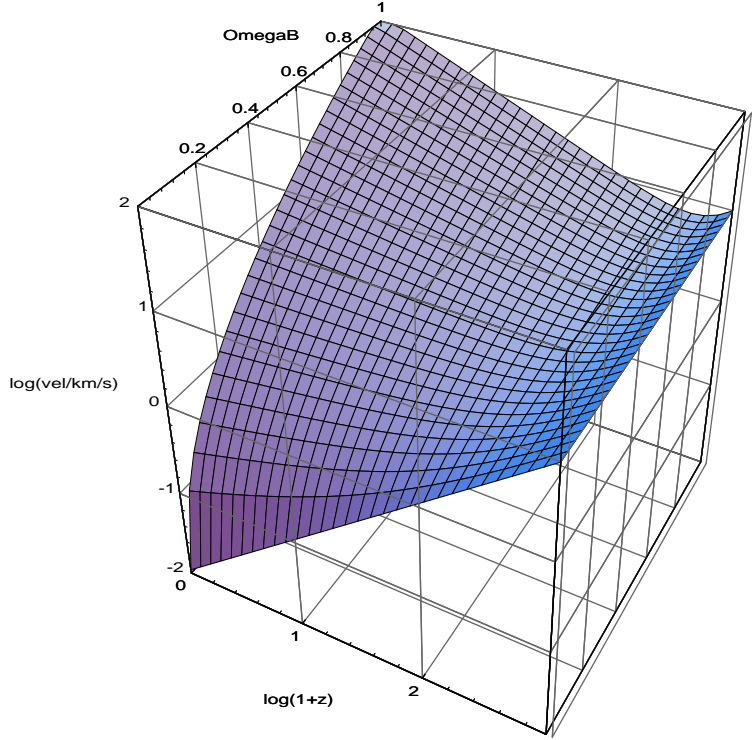,height=10cm,width=10cm}}
\caption[]{\small The peculiar velocity evolution for a string wake
  seeded at $z_{\rm kick}=900$ and $v_{\rm kick}=10 \kms$. The x--axis depicts
  the logarithm of redshift plus one, the y--axis the fraction of
  closure density in baryons, and the z--axis shows the logarithm of
  the peculiar velocity outside the shocked region.  Note that for
  $\Omega_B=0$ the velocity simply decays $\propto (1+z)$ due to the
  cosmological expansion. Note also that changing the initial velocity
  kick only shifts this surface along the z-axis but does not change
  its shape. }
\label{Fig:u}
\end{figure}

The trajectory of a gas particle after recombination and outside of the
shocked region is the same as for a collisionless particle, which for a
flat universe is given by (Stebbins \etal 1987)
\begin{eqnarray}
        \triangle_Q(z) = \frac{1}{(p_+ + p_-)} \frac{v_{\rm kick}}{H_0
          \sqrt{\Omega_0}} \left[ \frac{(1+z_{\rm kick})^{p_+ - 1/2}}
    {(1+z)^{p_+}} - \frac{(1+z)^{p_-} } {(1+z_{\rm kick})^{p_- + 1/2}} \right],
\end{eqnarray}
\noindent where
\begin{eqnarray}
  p_+ = \frac{1}{4} \left( \sqrt{1+ 24 \Omega_B/\Omega_0} -1 \right),
  \ \ \
  p_- = \frac{1}{4} \left( \sqrt{1+ 24 \Omega_B/\Omega_0} +1 \right)
\end{eqnarray}
where $\Omega_B$, $v_{\rm kick}$, and $z_{\rm kick}$ denote the fraction of
closure density in baryons, the initial kick velocity, and the kick redshift
respectively.  Since this result was derived using the Zel'dovich
approximation we know it is exact in one dimension.  Taking the time
derivative, we find for the evolution of the peculiar velocity of the gas
particles with respect to the cosmological frame, $u=(1+z)^{-1}
d{\triangle_Q}/dt$.  In Figure~\ref{Fig:u} we show the evolution of
the velocity perturbation induced by a string at redshift 900 as a function of
both redshift and $\Omega_B$ for the case of a flat background universe.
We see that for all cases with $\Omega_B\lsim0.5$, the infall velocity
will be smaller at the present time than the initial kick velocity,
illustrating that the wake was not able to acquire high enough surface
densities to overcome the ``drag'' from the Hubble expansion.

From the Rankine--Hugoniot relations (Rankine 1870, Hugoniot 1889) we
know that a strong adiabatic shock in an ideal gas will travel outwards
with a fraction $\frac{1}{2}(\gamma-1)$ ($=1/3$ for the mono--atomic
ideal gas with adiabatic index $\gamma=5/3$) of the infall velocity.
The maximum velocity difference at the shock is given by the peculiar
velocity $u$ with respect to the cosmological frame, yielding a maximum
post-shock temperature of $kT= \mu u^2/3$. The immediate post--shock
density equals $(\gamma+1)/(\gamma-1)$ ($=4$ for $\gamma=5/3$) times
the pre--shock density, $\rho_w = \rho_B(\gamma+1)/(\gamma-1) =
\rho_0 (1+z)^3 (\gamma+1)/(\gamma-1)$, where $\rho_0$ denotes the
present mean density of baryons in the universe.

Abel (1995) and Tegmark \etal (1997) showed how one can, for a
Lagrangian fluid element, integrate the kinetic rate equations of
molecular hydrogen formation for applications where the collisional
ionization of hydrogen atoms is unimportant, i.e.  low velocity shocks with
post--shock temperatures $\lsim 6000\K$.  For redshifts $\lsim 100$
they find the \HH fraction after one initial recombination time to be
given by
\begin{eqnarray}
        f_{H_2} = 1\tento{-3} \left( \frac{T_0}{2000\K} \right)^{1.53}
        \ln(x_0 n_H k_{\rm rec} t + 1) ,
\end{eqnarray}
where $T_0$ is the immediate post--shock temperature, $x_0$ the
initial free electron fraction, $k_{\rm rec}$ the hydrogen recombination
rate coefficient, and $n_H$ the number density of neutral hydrogen
atoms. This formula describes remarkably well the numerical results of Abel
\etal 1997b. For redshifts greater than 100, this solution can be used
as an estimate for the absolute maximum amount of \HH
that actually forms. We see
that it takes at least one initial recombination time
$(t_{\rm rec}=(k_{\rm rec} x_0 n_H)^{-1})$ to form a significant abundance of
\HH.  Using $x_0 = 1.2 \tento{-5} \sqrt{\Omega_0}/(h \Omega_B)$
(Peebles 1993), $k_{\rm rec} = 1.8\tento{-10} T^{-0.65} \cm^3 \s^{-1}$,
and the post--shock conditions derived above, we find that the minimum
time to form hydrogen molecules is given by
\begin{eqnarray}
  t_{\rm min} = 3.1\tento{15} \s\left(\frac{v_{\rm kick}}{10\kms} \right)^{1.3}
>
 \left(     \frac{1+z} {100} \right)^{-3}.
\end{eqnarray}
Hence we see that, contrary to the case of ionizing shocks, there is a
significant
time lag between when the shock passes by and when enough \HH
can be formed to enable
the gas to cool appreciably.  At redshifts greater than 100, this time
delay is longer due the photo-dissociation of the intermediaries \Hm
and \HHp by the CMBR. This is essentially the reason why the instability
of radiative shocks, as discussed by Chevalier and Immamura (1982), is not
found for primordial gas when the shock does not ionize the gas.
However, this instability can be observed for models with $\Omega_B >
0.5$ since these models will be able to produce shocks with velocities
faster than the initial kick
velocity (see Figure~\ref{Fig:u}) and reach the part of the cooling
curve that allows the instability.

\section{NUMERICAL SIMULATIONS}
\label{sec:simulations}

\subsection{METHODS AND INITIALIZATION PARAMETERS}

A numerical code with high spatial and  mass resolution is required to
model the hydrodynamics of a cosmic string wake  and the micro--physics of
chemical   reactions   and radiative   cooling.  We  can  achieve high
dynamical  ranges  with the two--dimensional  hydrodynamics code of Anninos
and  Norman (1996) which includes the  9 species chemistry and cooling
model  of  Abel \etal  (1997)  and  Anninos \etal  (1997).
However, we have replaced
the ground state \HHp photo-dissociation  rate of Abel \etal
(1997) with the LTE  rate using the equilibrium constant given
by Sauval  \& Tatum (1984) to account  for  the close coupling between
the CMBR and the \HHp molecule at high redshifts.
We also  account for the  fact that the  baryonic fluid motion at high
redshift is slowed  by the scattering of CMBR photons  off the residual free
electrons, the so called Compton Drag. This additional force can be
expressed by a ``drag time'', $t_\rmd=v/\dot{v}$, where $v$ denotes
the peculiar velocity of the free electrons.
For   a  gas  with   fractional ionization $x_\rme=n_\rme/n_\rmt$  and
density $\rho_\rmb = n_\rmt \mu m_\rmp$,
where $n_\rme$ is the free electron number density,
$n_\rmt$ is the total particle
number density and $\mu$ is the mean molecular  weight in units of proton
mass $m_\rmp$, one may write this ``drag time'' as
\begin{equation}
\label{CD3}
t_\rmd={3\over4}\,
 {1\over x_\rme}\,{\mu\, m_\rmp\over\sigma_\rmT\,c\,\rho_\gamma}
        =2.4\times10^5\,{\mu \over x_\rme}\,
        \left({1000\,\rmK\over T_\gamma}\right)^4 \yrs.
\end{equation}
where  $c$  is  the speed of   light,   $\sigma_\rmT$ is  the  Thomson
cross--section, and  $T_\gamma$ is  the  CMBR temperature  in  Kelvin.
\Eq{CD3}  is  only valid in   the optically thin   limit,  since in an
optically  thick medium  the photons and  baryons act  as one fluid on
scales much greater than the  photon mean free  path.  In the code, we
simply  modify  the  updated  velocity according to   the  first order
discretization $\Delta  \bfv   = -  \bf{v}  \Delta  t/t_{\rmd}$, where
$\Delta t$ is the timestep defined by the  minimum of the hydrodynamic
Courant,  gravitational  free--fall, and cosmological expansion times.
This will be sufficiently  accurate  as long  as the ``drag  time'' is
much longer than the Courant time step.   We start all our simulations
at redshift $z=1100$ with an  initial \HH  fraction of $10^{-10}$  and
evolve the uniform background to the ``kick redshift''.  For the other
eight species, we use the same initialization as in Anninos and Norman
(1996).   Furthermore, this procedure ensures  that the temperature at
the kick  redshift  $z_{\rm  kick}$,  is  correctly derived   from the
balance of adiabatic expansion cooling and Compton heating.

An infinitely  long and straight  string traveling with a speed $v_S$
causes a velocity boost $v_{\rm kick}$  towards the sheet given by
(Stebbins~\etal~1987, but see \S\ref{sec:Wiggly Strings})
\begin{equation}
  \label{vkick}
  |v_{\rm kick}| =  4 \pi \frac{G \mu}{c^2} v_S \left[ 1 - \left( \frac{v_S}{c}
    \right)^2 \right]^{-1/2}  ,
\end{equation}
where $G$ is the gravitational constant and
$\mu$  is a mass per unit length of the string.
Measurements of the CMBR anisotropy by the COBE satellite indicate
that $G\mu/c^2\simlt1.5\times 10^{-6}$ (Bennett, Stebbins, \&
Bouchet 1992, Coulson \etal 1994, Allen \etal 1996). Hence, for typical string
velocities of $v_s\sim 0.8c$, one finds $v_{\rm kick}\sim6\kms$.

To simulate   the idealized  situation of  a   perfectly  straight and
infinite  string,    it is   sufficient  to  perform  one--dimensional
simulations since    the  sheet traced  out by    the moving string is
translation symmetric.  Throughout this study we use a logarithmically
refined grid spanning a physical distance of $0.4$ comoving \Mpc, with
200 zones arranged such that the  zone size decreases from $18$\kpc at
the outermost  zones  to  $1.4$\pc at    the central  zone   along the
collapsing  direction.  The    neighboring   cells have    a  constant
refinement  ratio  of  $r=1.1$.  We  also set   $\Omega{HDM}  = 0.94$,
$\Omega_B=0.06$ and $h=0.5$ in all the simulations.

\subsection{WIGGLINESS}
\label{sec:Wiggly Strings}

It has  been known  for some time  that  cosmic string networks, while
generally having a  coherence  length close to the cosmological  horizon,
also have structure on scales  much smaller than the horizon (Vilenkin
\&  Shellard 1995,  and references  therein).   These  wiggles do  not
extend to arbitrarily small scales, as they are damped by the emission
of gravitational  radiation.   While  the rms speed  of  cosmic string
segments must be close to the speed of light, the ``bulk velocity'' of
a long  length of  string, including the  oscillating wiggles,  can be
significantly less.  A particle located  at a distance from the wiggly
string that is greater than the extent of the wiggles would experience
a  net gravitational attraction  greater than indicated by \eq{vkick}.
However if one is interested  in the  gas  dynamics in a small  region
around a segment of the string, a region smaller than the scale of the
wiggles,  the gravitational forces will  be dominated  by the force of
that  one segment  and \eq{vkick} will  give a  good indication of the
velocity perturbation of the passing string.

Now we   determine which regime   we  should be  working  in.   For  a
non-wiggly  string, $v_{\rm kick}\sim 10  G\mu/c$  and the velocity of
the strong shock wave produced  would be ${1\over3}$  of this, so  the
comoving width of the baryonic wakes is $\sim 5 (1+z)\,G\mu t/c$.  The
comoving gravitational back-reaction scale  for the strings, the scale
of the  smallest wiggles, is estimated  to be $\sim100(1+z)\,G\mu t/c$
(Vilenkin \& Shellard 1995, and reference therein).  While the two are
the same order in $G\mu$,  the numerical factors  point to the wiggles
being  larger than the width  of the baryonic  wakes.  If  this is the
case, then a purely planar wake might provide a sufficient description
for the  wakes.  However, the  effects of  gravitational back-reaction
are not completely understood and there may be other numerical factors
which  might cause these  two scales to  be more comparable.  We allow
for this in   two  ways in   our simulations: first  we  consider kick
velocities up   to 20km/sec,  which is greater    than  what would  be
expected for a straight piece of string, but might be more typical for
the net effect of a wiggly string; secondly,  in our 2--d simulations,
we   consider  the  flow   onto    a non-uniform string   world-sheet,
effectively modulating the value of $v_{\rm kick}$.  More specifically
we have modeled the initial  flow as a  potential flow, and one  whose
divergence  $\nabla\cdot\bfv_{\rm  kick}$  is non-zero on  the  planar
string world-sheet, just as in the 1-d  case.  However in the 2-d case
we allow the   divergence to vary  in  one direction  along  the world
sheet.  More  specifically, this divergence  is drawn  from a random
distribution given by the positive square  root of the sum
of  the square  of  two identically distributed  real  Gaussian random
noise variables.  From the   perturbed surface  density  we compute the
spatially varying  velocity perturbations which show reflection symmetry
about the string worldsheet.  The variation along the   wake was  numerically
captured with up to 128 grid points and the box size in this direction
varied    from   640 to  32 comoving    kpc.    We note, however, that
introducing a spatially non-uniform  second dimension does not lead to
significantly higher overdensities than in the 1-d case.

\subsection{RESULTS}
\label{sec:results}

We investigate wakes from infinitely long and fast moving straight strings
with different initial data:
$v_{\rm kick} =  20$, 15, 10, $5\kms$, and
$z_{\rm kick}= 900$, 500, 100,  and 50.  We discuss  here only the results
from one  particular one--dimensional simulation,  with $z_{\rm kick}=900$
and $v_{\rm kick}=20 \kms$. This optimistic case of an early string with a
very high kick  velocity is the most  promising candidate to  form the
first structures in the cosmic string + HDM model and, indeed, led to the
highest over densities ($\sim 200$) in all the simulated cases.

\begin{figure}[ht]
\centerline{\psfig{file=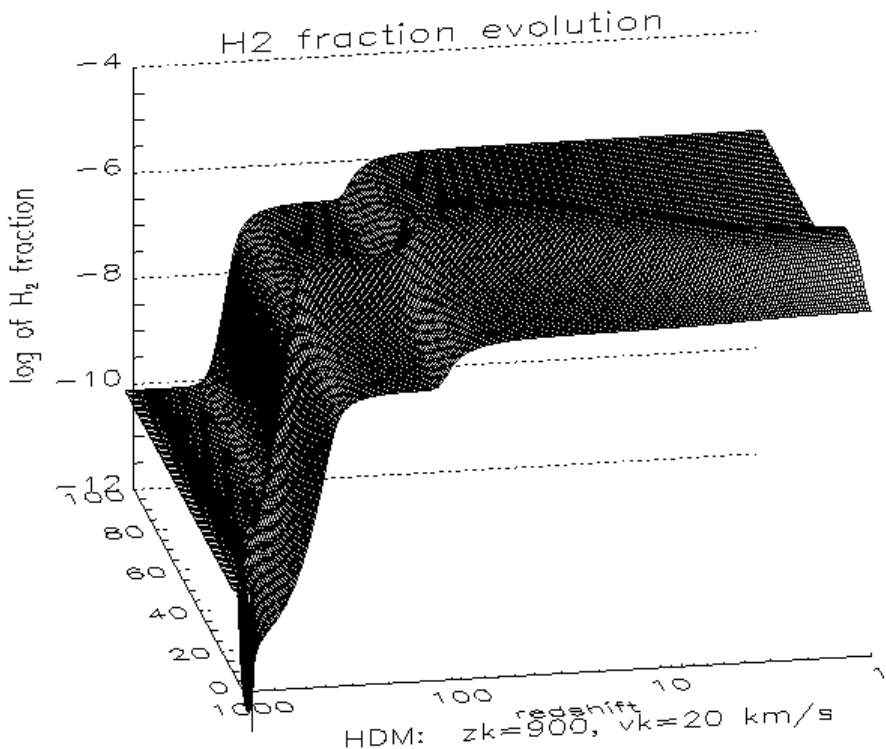,height=10cm,width=12cm}}
\caption[]{\small The evolution of the  \HH fraction for a string with
  $z_{\rm kick}=900$ and $v_{\rm kick}=20  \kms$.  The x--axis depicts
  the redshift, the y--axis the cell number in the 1--d simulation and
  the z--axis shows the logarithm of the \HH fraction.}
\label{Fig:fH2900v20}
\end{figure}

Figure \ref{Fig:fH2900v20} shows,    for  one side of  the   perfectly
symmetric wake, the  evolution   of  the  mass fraction of    hydrogen
molecules vs.  cell number (distance from  the string worldsheet), and
also  vs. redshift on  the x--axis.   At the kick  redshift $z=900$, a
shock forms  that runs outwards.  Two distinct  steps are evident: the
first at redshift  $z\sim 500$, and the second  at $z  \sim 120$.  The
plateau  before the  first   step  is an  artifact  of  our  too  high
initialization of  \HH and is not physical.   This,  however, does not
influence the subsequent evolution and results.  The first step is due
to the \HHp chemistry, and occurs when  the formation of \HH molecules
starts  to   dominate   the destruction   of   \HHp  compared  to  its
photo--dissociation by the CMBR. In the  shocked region, it is evident
that the   $v_{\rm   kick}=20 \kms$ shock    is  initially capable  of
destroying some  hydrogen molecules.   Shortly  after $z_{\rm   kick}=
900$, however, the decaying  infall velocity, the steadily rising \HHp
abundance, and  the increasing \HH  formation rate lead to  an overall
enhanced \HH fraction.  The second step is due  to the \Hm path, which
occurs  slightly earlier in  the  denser, shocked regions  than in the
background.  This is  easily understood from the equilibrium abundance
of \Hm which is  given by $n_{H^-} =  n_e n_H k_7/(k_{23} + k_8 n_H)$,
where  $k_7$,  $k_8$, and $k_{23}$  denote  the reaction rates for the
photo--attachment of \Hm, the dissociative  attachment reaction of \Hm
(\HH   formation),  and the  photo--detachment   by  the CMBR of  \Hm,
respectively (the notation is taken from Abel \etal 1996). That second
step originates when $k_{23} \gsim  k_8 n_H$, and $n_{H^-}$ is roughly
proportional to $n_H  n_e$ which is higher in  the shocked gas than in
the background, yielding to the enhanced \HH formation rate.  Another,
and at first  surprising, effect is  that  the central regions  of the
wake form fewer hydrogen molecules than regions further away.  This is
because the gas closer to the  string world--sheet shocks at redshifts
$z>110$, at  which \Hm is still destroyed  efficiently by the CMBR and
\HH cannot  form.  The higher  density in this shocked component leads
to  an   increased hydrogen recombination     which leaves fewer  free
electrons    as    catalysts     for  \HH     formation    at    lower
redshifts. Consequently, this \HH depression in the central layers are
not observed for simulations with kick redshifts $\lsim 110$.  

Compton Drag never becomes dynamically important even in the cases for
which we chose extremely high kick velocities for  which the shock was
ionizing   the  gas.  This   is because   the  wakes  do   not  become
self--gravitating until very late redshifts.  In other words, the post
shock gas continues  to expand with the  Hubble flow and the gas shows
no   motion with   respect  to  the    rest frame   of  the background
photons. Hence, there will be no net force  on the baryons independent
of their ionized  fraction.  Since the self-gravity  of the  gas plays
little  role in the  evolution of the wake  at  early times, we should
expect   that inhomogeneities in the  wake  should  have little effect
since there is no significant gravitational instability to amplify any
initial inhomogeneities.

Shapiro and Kang (1987) presented calculations of steady--state shocks
of 20,  30, and  50\kms\ occurring at  redshifts  100 and 20 with  and
without external radiation fields. Interpreting their results as upper
limits   on the  possible  \HH  formation  and   cooling  we find  our
simulations to agree with their findings at the low redshifts and high
kick velocities where they are comparable.

\section{CONCLUSION}
\label{sec:discussion}

We have  studied the possibility  of high redshift structure formation
in the wakes of fast,  long, and straight, as  well as fast, long, and
wiggly cosmic strings in a hot dark matter dominated universe. We find
that  insufficient amounts of \HH   molecules are formed, and the  gas
does not cool  appreciably.  Due to  the fact that the baryonic cosmic
string wakes do  not  become significantly self--gravitating to  allow
the infall velocities to exceed the initial  kick velocity (see Figure
1.), we find the maximum overdensities  reached in all the simulations
we  have carried  out are $<  10^3$, corresponding  to hydrogen number
densities  less than   $\sim 1\tento{-4} (\Omega_Bh^2/0.025)  (1+z)^3
\cm^{-3}$.  Thus no   primordial stars can  be formed   in the studied
wakes.

	This leads  us  to conclude that, in  the  cosmic string + HDM
model, one  must wait  for  the neutrinos to collapse  before anything
like  star formation can take  place. In this  work we have explicitly
considered  gauge  (local)   cosmic  strings,  but   we  expect  these
conclusions will apply to global cosmic strings  as well. The baryonic
structures considered above will be  destroyed if and when they become
enveloped in the collapse of the  neutrinos, which will happen on much
larger scales  and  involve much   larger   velocities.  It  has  been
suggested that neutrino  collapse may happen   as early as  $z\sim100$
(Zanchin,   Lima, \& Brandenberger  1996;   Sornborger 1997) but other
studies     (Colombi  1993)    find  non-linear    collapse  only  for
$z\simlt5$. Thus it is still an unsettled issue whether cosmic strings
could have  brought light   to the  dark ages,  at  least in   the HDM
scenario.

\acknowledgements

We are  grateful  for  discussions with Andrew  Sornborger  and Robert
Brandenberger.  This  work  is done under   the auspices of  the Grand
Challenge  Cosmology Consortium  (GC3) and supported   in  part by NSF
grant ASC--9318185.  The  simulations were performed  on the CRAY--C90
at the Pittsburgh Supercomputing center.

\end{document}